\documentclass[fleqn,twoside]{article}
\usepackage[headings]{espcrc2}
\usepackage{epsfig}
\readRCS
$Id: espcrc2.tex,v 1.2 2004/02/24 11:22:11 spepping Exp $
\usepackage{graphicx}
\usepackage[figuresright]{rotating}

\newcommand{\AmS}{{\protect\the\textfont2
  A\kern-.1667em\lower.5ex\hbox{M}\kern-.125emS}}

\title{New Physics Search in Flavour Physics
\thanks{CERN-PH-TH/2005-168, SLAC-PUB-11530; based on invited talks given 
at {\it Beauty 2005}, the 
10th International Conference on $B$-physics  at  Hadron Colliders, June 20-24, Assisi (Perugia, Italy)  and 
at {\it QCD@Work}, International Workshop on Quantum Chromodynamics, 
June 16-20, Conversano (Bari, Italy)}}

\author{Tobias Hurth\address{CERN, Dept.\ of Physics, Theory Unit, CH-1211 Geneva 23, Switzerland\\ SLAC, Stanford University, Stanford, CA 94309, USA}
\thanks{Heisenberg Fellow}}

\runtitle{New physics search in flavour physics}
\runauthor{Tobias Hurth}

\begin{document}

\begin{abstract}
With the running $B$, kaon and neutrino physics 
experiments,  flavour physics takes centre stage within today's 
particle physics.  We discuss the opportunities offered by these 
experiments in our search 
for new physics beyond the SM   
and discuss  their  complementarity to collider physics. 

We focus on rare $B$ and kaon decays,
highlighting specific observables in an exemplary mode. 
We also comment on the so-called $B \rightarrow \pi\pi$ 
and $B \rightarrow K \pi$ puzzles. 
Moreover, we briefly  discuss the 
restrictive role  of long-distance strong interactions and some 
new tools such as  QCD factorization and SCET to handle them.
\end{abstract}

\maketitle

\section{Introduction}
There are three main issues  within the present $B$ and kaon physics
programme: 
(i) the search for new degrees of freedom beyond the standard model  (SM) 
in flavour-
or CP-violating processes; (ii) the question of the precise mechanism 
of CP violation and (iii) the search for a quantitative understanding of 
the strong interactions within flavour observables.

Rare $B$ and kaon decays representing  loop-induced processes 
are highly sensitive  probes for new degrees of freedom  beyond the SM.
Through  virtual (loop) contributions of new particles  
to such observables, one can investigate  high-energy scales even before 
such energies are accessible at  collider experiments. 

Such  flavour information is complementary to the collider data
of the Large Hadron Collider (LHC); 
for example the present flavour data from the $B$
factories in the $b \rightarrow s$  sector 
is not very restrictive (yet). Within a supersymmetric new physics scenario,  
present bounds on squark mixing still allow for large contributions to
flavour-violating squark decays at tree level,  which can be measured 
at the LHC.  In these cases, additional information from 
future flavour experiments will be necessary  to interpret
those LHC data properly.

The CKM prescription of CP violation with one single phase is very
predictive. Before the start of the $B$ factories, the neutral kaon 
system was the
only environment where CP violation had been observed. It was 
difficult to decide if the CKM description of the SM
really accounted quantitatively for the CP violation observed in the
kaon system, because of the large theoretical uncertainties due to
long-range strong interactions. The rich data sets from the
present and planned $B$ experiments  now allow for an independent 
and really quantitative test of the CKM-induced CP-violating effects 
in several independent channels.

Quark-flavour physics is governed by the interplay of strong and 
weak interactions.  One of the main difficulties in examining the 
observables in  flavour physics is the influence of the 
 long-distance strong interaction. 
The resulting hadronic uncertainties restrict the opportunities in
flavour physics significantly.
If new physics does not show up in $B$ physics through large deviations, 
as  recent experimental data indicate, the focus on theoretically clean 
variables within the indirect search for new physics  is mandatory.

Nevertheless there are new tools, such as  QCD factorization 
and the soft-collinear effective theory (SCET),  to tackle 
the strong interaction within $B$ decays. 
The large data sets from the $B$ experiments should 
be used to sharpen these new tools and improve our present 
understanding of the strong interaction.

\subsection{Experimental roadmap}

The present experimental roadmap of flavour physics  
offers great  opportunities.  
Several $B$-physics experiments are successfully
running at  the moment and, in the upcoming years, 
new facilities will start
to explore  $B$ physics with increasing 
sensitivity and within various    
experimental settings. 
There are two $B$ factories,  operating at the  $\Upsilon (4S)$ 
resonance in an asymmetric mode, successfully obtaining data, namely 
the BABAR experiment at SLAC (Stanford, USA) \cite{Babar}  
and the BELLE experiment  at KEK (Tsukuba, Japan) \cite{Belle}. 
An upgrade of the BELLE machine is planned; and there 
are also plans for Super-BELLE \cite{Superbelle}. 
After the present hadronic 
$B$-physics programme  at FERMILAB (Batavia, USA) \cite{Tevatron},  
there are  strong $B$-physics programmes  planned at  three 
LHC experiments  at CERN in Geneva, especially at LHCb \cite{LHCb}. 
There is also a future option of a $B$-physics programme at 
a future linear collider via a  Giga-Z factory (see \cite{GigaZ}). 

The main motivation for a $B$ physics programme  
at hadron colliders  is the huge $b$-quark production cross section 
with  respect to  the one at 
$e^+ e^-$ machines, and the opportunity to analyse  also the $B_s$ system. 
Nevertheless, a future Super-$B$ factory would be  
competitive but also complementary to the planned hadronic $B$ physics 
programme (see for details \cite{Superbelle2,Superbabar}). 

There are many further sectors of flavour physics 
 that  offer important experimental opportunities. 
$K$ decays such as  $K \rightarrow \pi \nu \bar\nu$ and
$K_L \rightarrow \pi^0 \ell^+ \ell^-$ are extremely sensitive to
possible new degrees of freedom and are largely  unexplored.  In fact,
at present we have fewer  constraints on short-distance-dominated 
$s \rightarrow d$ quark transitions than on $b \rightarrow s$ ones.
In the presence of new physics, charm physics could provide
important inputs by future 
$e^+e^-$ and fixed-target experiments.
Searches for electric dipole moments of various  
particles  are  a very  important  source  of information on the 
flavour and CP structure.  
Open questions in neutrino physics, regarding their
masses, their mixing and their particle nature, are actively being
attacked  in the present and future  experimental programme. 
The study of the correlation of neutrino properties with flavour 
phenomena in the charged-lepton and in the quark sector, e.g.  
 charged-lepton flavour violation, 
is also an important target. Pushing the present limits 
on $\mu \leftrightarrow e$
and $\mu \leftrightarrow \tau$ transitions might lead to important 
insight.

\subsection{Hadronic uncertainties}

The crucial problem in the new physics search within flavour physics is 
the optimal separation of new physics effects and hadronic uncertainties.
This can be successfully solved only for a selected number of 
{\it golden} observables in flavour physics,  
where hadronic physics can be disentangled to a large extent 
and clean tests of the SM  are possible. 
In principle there are three 
strategies:

$\bullet$ One can focus on inclusive decays modes. These modes 
are  dominated by the partonic contributions 
because  bound-state effects of the final states are eliminated
by averaging over a specific sum of hadronic states.  Moreover, also
long-distance effects of the initial state are accounted for, through
the heavy mass expansion in which the inclusive decay rate of a heavy
$B$ meson is calculated, using an expansion in inverse powers of the
$b$-quark mass. In fact, one can use
quark-hadron duality to derive a heavy mass
expansion of the decay rates in powers of $\Lambda_{\hbox{\tiny
    QCD}}/m_b$ (HME). For example, it turns out that the decay
width of the $\bar B \rightarrow X_s \gamma$ is well approximated by the
partonic decay rate, which can be calculated in
renormalization-group-improved perturbation theory:
\begin{eqnarray} \nonumber
\Gamma (\bar  B \rightarrow X_s \gamma) = \Gamma ( b \rightarrow X_s^{parton} 
\gamma ) + \Delta^{nonpert.}  
\nonumber
\end{eqnarray} 
Non-perturbative corrections  
occur at the  order  $\Lambda^2_{QCD}/m_b^2$ only. The absence of first-order 
power corrections 
is a consequence of the fact that there is no independent 
gauge-invariant operator of dimension 4 in the operator product expansion 
because of the equations of motion. 
The latter fact
implies a rather small numerical impact of the non-perturbative
corrections to the decay rate of inclusive modes.
Nevertheless, there are additional nonperturbative corrections within
\mbox{inclusive} modes due to necessary cuts in the experimental spectra like
the photon energy spectrum in $\bar B \rightarrow X_s \gamma$ (see \cite{Neubertnew}).

$\bullet$  In exclusive processes, however, one cannot rely on 
quark-hadron duality and has to
face the difficult task of estimating matrix elements between meson
states. Therefore, exclusive modes are not well-suited to the new physics search in general. Nevertheless, one can focus on ratios of exclusive decay 
modes such as asymmetries, where large parts of the hadronic uncertainties partially cancel out. In particular, there are CP asymmetries that are governed
by one weak phase only. In that specific case the hadronic matrix elements
cancel out completely.

$\bullet$ There are also specific decays like 
$K \rightarrow \pi \nu \bar\nu$ modes where the hadronic 
uncertainties can be eliminated by experimental data. In 
these kaon decays the hadronic matrix element can be related 
to the well-known rare semileptonic $K_{l3}$ decays.

Regarding the hadronic matrix elements of exclusive modes,  
the method of QCD-improved factorization has been 
systemized for non-leptonic decays in the heavy-quark limit. This method 
allows for a perturbative calculation of QCD corrections to naive 
factorization and is the basis for the up-to-date predictions for exclusive 
rare $B$ decays in general \cite{Beneke:1999br}. However,
within this approach, a general, quantitative method to estimate the
important $1/m_b$ corrections to the heavy-quark limit is missing.

A  more general quantum field theoretical framework was proposed 
-- known under the name of
SCET -- which allows for a deeper 
understanding of the QCD factorization approach \cite{SCETa,SCETb}.
 In contrast to the well-known 
heavy-quark effective theory (HQET), the recently 
proposed SCET does not correspond to a local operator expansion.
While HQET is only  applicable to $B$ decays, when the energy transfer 
to light hadrons  is small, for example to  $B \rightarrow D$ transitions 
at small recoil to the $D$ meson, it is not applicable, when 
 some of the outgoing, light particles have momenta of order $m_b$; 
then one faces a multi scale problem:\\
a) $\Lambda = {\rm few} \times \Lambda_{\rm QCD}$,
the {\it soft} scale set by the typical energies and
momenta of the light \mbox{degrees} of freedom in the hadronic bound states;
b) $m_b$ the {\it hard}\/ scale set by the heavy-$b$\/-quark mass (we note, 
that in the $B$\/-meson rest frame, for $q^2\simeq 0$
also the energy of the final-state hadron is 
given by $E\simeq m_b/2$);
c) the hard-collinear scale $\mu_{\rm hc}=\sqrt{m_b
\Lambda}$ appears through interactions between soft and energetic
modes in the initial and final states. The dynamics of hard and hard-collinear
modes can be described perturbatively in the heavy-quark limit 
$m_b \to \infty$.

The separation of the two perturbative scales 
from the non-perturbative hadronic dynamics is formalized, 
within the framework of SCET,  with the small expansion parameter 
$\lambda = \sqrt{\Lambda/m_b}$.
Thus, SCET describes $B$ decays to light hadrons with energies 
much larger than their masses, assuming that their constituents have 
momenta collinear to the hadron momentum. 
On a technical level, the implementation of power counting 
in $\lambda$,  at the level  of momenta, field and operators,  
corresponds directly   to the
well-known method of regions for Feynman diagrams \cite{Regions}.

The large varity of experimental data on those decay modes allows  
us to test these  new  tools  
and perhaps to reach sufficient accurary for the determination 
of the CKM parameters and even for the detection of new physics effects.

\section{Exploration of higher scales via rare decays}
\begin{figure}[htbp]
\begin{center}
\vspace{-0.5cm}
\includegraphics[scale=0.42]{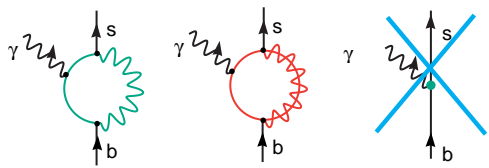}
 \vspace{-1cm}
\caption{Standard and nonstandard contributions to $b\rightarrow s \gamma$ at the-loop level only \label{indirect}.}
\end{center}
\end{figure}

Rare $B$ and kaon processes often 
represent flavour changing neutral currents (FCNCs)
and occur in the SM  only at the loop  level. 
This fact leads to the high sensitivity to potential new degrees of freedom 
beyond the SM.
Such potential new contributions are not suppressed  
with respect to the SM  contributions (see Fig.\ref{indirect}). 
This  indirect search for new physics 
signatures  within flavour physics takes place today in complete darkness, 
given that  we have no direct evidence of new particles beyond the SM.  
However, the day the existence of new degrees of freedom  is established  by 
the LHC, the searches for anomalous  phenomena  in the flavour sector will  
become mandatory.  The problem  then will no longer be to discover new 
physics,  but to measure its (flavour) properties.  In this context, 
the measurement of  theoretically  clean rare  decays, {\it even when found 
to be SM-like}, will lead to  important and  valuable  information  of 
the structure of the new physics models and will lead to complementary
information to the LHC collider data. 

Because  new physics effects beyond the SM seem to be rather small
it is important to go beyond 
the pure study of branching ratios and also look at complex kinematic 
distributions, in particular at CP, forward-backward, isospin and polarization 
asymmetries. Only the measurements of a large overconstraining 
set of these observables allow us to detect specific pattern and to distinguish
between various new-physics scenarios.

Finally, rare decays are also important tools to analyse the famous 
flavour problem, namely how FCNCs  are 
suppressed beyond the SM. This problem has to be solved by any viable
new physics model. One solution of the flavour problem is given by 
 minimal flavour violation (MFV).
In \cite{Giannew}, a consistent definition of this scenario was
presented, which essentially requires that all flavour and CP-violating 
interactions be linked to the known structure of Yukawa couplings. 
The constraint within an  effective field approach  is introduced  
with the help of a symmetry concept and can be shown to be 
renormalization-group-invariant \cite{Giannew}.

Perhaps this MFV-based effective field theory approach is too 
pessimistic from the current  point of view. 
One of the key predictions of the MFV is the direct link 
between the  $b \rightarrow s$, $b\rightarrow d$,  and $s \rightarrow d$
transitions. This prediction within the $\Delta F = 1$ sector is definitely 
not well-tested at the moment and  there is still room for new flavour 
structures to be discovered. 
Nevertheless, in contrast to the scale of the electroweak
symmetry breaking, there is no similarly strong argument 
that new flavour structures have to appear at the electroweak scale.

\subsection{$b \rightarrow s/d \gamma$ and $b \rightarrow s \ell^+\ell^-$ modes}

The inclusive $b \rightarrow s \gamma$ mode is still the most prominent
rare decay, 
because it has already measured by several independent experiments
\cite{aleph,belle,cleobsg,babar1,babar2,Koppenburg:2004fz} 
and the present experimental accuracy has reached the $10 \%$ level
\cite{Alexander:2005cx}:
\begin{eqnarray}
\mbox{BR}[\bar B \to X_s \gamma] = (3.52 \pm 0.30) \times 10^{-4} \,.
\label{world} \nonumber
\end{eqnarray}
In the near future, more precise data on this mode are expected from the $B$
factories. Thus, it is mandatory to reduce the present theoretical uncertainty 
accordingly. A systematic improvement certainly consists in performing a
complete NNLL calculation which will reduce the well-known large 
uncertainty due to the definition of the charm mass \cite{Gambino} 
by a factor 2 as was recently shown \cite{New}.   
In a recent theoretical update of the NLL prediction of this branching ratio,
the uncertainty related to the definition of $m_c$ was taken into account
by varying $m_c/m_b$ in the conservative range $0.18 \le m_c/m_b \le 0.31$
which covers both, the pole mass (with its numerical error) value and 
the running mass
$\bar{m}_c(\mu_c)$ value with $\mu_c \in [m_c,m_b]$ \cite{Hurth:2003dk}:
\label{hurth_lunghi}
$\mbox{BR}[\bar{B} \to X_s \gamma] = (3.70 \pm 0.35|_{m_c/m_b} \pm
0.02|_{\rm{CKM}} \pm 0.25|_{\rm{param.}} \pm 0.15|_{\rm{scale}}) \times 10^{-4} \, .$
The stringent bounds obtained from the $B \rightarrow X_s \gamma$ mode
on various non-standard scenarios (see
e.g.~\cite{Degrassi:2000qf,Carena:2000uj,Giannew,Borzumati:1999qt,Besmer:2001cj,Ciuchini:2002uv} are a clear example
of the importance of clean FCNC observables in discriminating
new-physics models.  

\begin{figure}[!t]
\hbox to\hsize{\hss
\includegraphics[width=0.85\hsize]{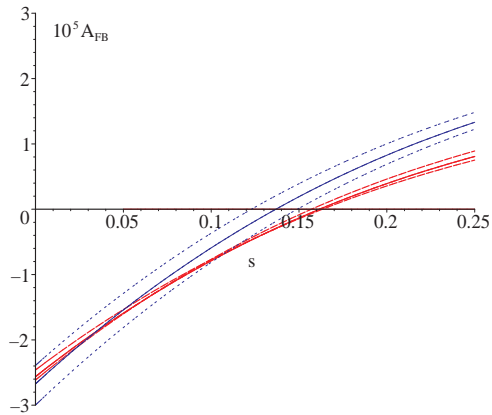}
\hss}
\caption[1]{Comparison between NNLL and  NLL results for 
$A_{FB}(s)$ in the low $s$ region. 
The three thick lines are the NNLL  
predictions for $\mu=5$~GeV (full),
and $\mu=2.5$ and 10 GeV (dashed); the dotted  curves 
are the corresponding NLL results. All curves for $m_c/m_b=0.29$.}
\label{fig:AFB} 
\end{figure}

Besides the $b \rightarrow s \gamma$ mode, also the $b \rightarrow s
\ell^+\ell^-$ transitions are already accessible at the $B$ factories
\cite{BELLEbsll1,BABARbsll1,BELLEbsll2}, inclusively and exclusively.
Quite recently also the $b \rightarrow d \gamma$ transition was 
measured for the first time \cite{bd} (for a recent review see
\cite{Hurth:2003ej}). 
The inclusive decay $b \rightarrow s \ell^+\ell^-$ is particularly 
attractive because of kinematic observables such as 
the invariant dilepton mass spectrum and the forward--backward 
(FB) asymmetry. 
This inclusive decay  is  also dominated 
by perturbative  contributions if the  $c \bar c$ resonances 
that show up as large peaks in the dilepton invariant mass 
spectrum are removed by appropriate  kinematic cuts. 
In the 'perturbative windows', namely  
in the low-${s}$ (dilepton mass) \mbox{region} 
($ 0.05 <  {s} = q^2 / m_b^2 < 0.25 $ and 
also in the high-${s}$ region with $ 0.65 < {s}$,
a theoretical precision comparable with  the one reached  
in the decay $b \rightarrow s \gamma$ is in principle possible. 
The recently calculated NNLL contributions 
\cite{Asa1,Adrian2,Adrian1,Asa2,MISIAKBOBETH,Gambinonew}
have significantly 
improved the sensitivity of the inclusive $\bar B \rightarrow X_s \ell^+ \ell^-$ 
decay in  testing extensions of the SM in the sector of flavour 
dynamics, in particular the value of the dilepton invariant mass
($q^2_0$), for which the differential forward--backward \mbox{asymmetry}
vanishes, is one of the precise predictions in flavour physics
 (see Fig.\ref{fig:AFB}): 
\begin{eqnarray}  q^2_{0,\,{\rm NNLL}} = (3.90 \pm 0.25)~\hbox{GeV}^2.
\label{eq:s0NLO} \nonumber
\end{eqnarray}
Let us briefly comment on the 
 impact of the exclusive
$B\to K^{(*)} \ell^+\ell^-$ modes. Hadronic uncertainties on 
these exclusive rates are dominated by the
errors on form factors and are much larger than in the corresponding
inclusive decays. In fact, following the analysis presented in
Ref.~\cite{AGHL}, we see that inclusive modes already put much
stronger constraints on the various Wilson coefficients. 
Concerning the measurement of a zero in
the spectrum of the forward-backward \mbox{asymmetry}, things are different.  
According to
Refs.~\cite{BeFeSe} the value of the dilepton invariant mass
($q^2_0$), for which the differential forward--backward asymmetry
vanishes, can be predicted in quite a clean way. In the QCD
factorization approach at leading order in $\Lambda_{\hbox{\tiny
    QCD}}/m_b$, the value of $q_0^2$ is free from hadronic
uncertainties at order $\alpha_s^0$ (a dependence on the soft form
factor $\xi_\perp$ and the light cone wave functions of the $B$ and
$K^*$ mesons appear at NLL). Within the SM, the authors of
Ref.~\cite{BeFeSe} find: $q_0^2 = (4.2 \pm 0.6) \, \hbox{GeV}^2$. 
As in the inclusive case, 
such a 
measurement will have a huge phenomenological impact.


\subsection{There is also beauty in kaon physics}

Although the general focus within flavour physics is 
at present on $B$ systems, kaon physics 
offers interesting complementary opportunities in the new 
physics search,  such as the exclusive rare
decays $K^+ \rightarrow \pi^+ \nu \bar{\nu}$ and 
 $K_L \rightarrow \pi^0 \nu \bar{\nu}$. 
These decay modes are extremely sensitive to possible new degrees 
of freedom, but they also allow for an accurate determination of the unitarity
triangle, which is completely independent from that of  the $B$ 
system (for a recent review, see \cite{Buras:2004uu}).

These modes are basically unexplored yet.
While there is only an upper limit on the neutral mode, three events 
were found in the charged mode by the AGS E787 and the E949 Collaborations
at Brookhaven \cite{Adler:2004hp,Anisimovsky:2004hr}, leading to 
\begin{eqnarray} \nonumber
{\rm BR}(K^+\to\pi^+ \nu\bar{\nu}) = \left( 1.47^{~+~1.3}_{~-~0.9} \right) \times 10^{-10}.
\end{eqnarray} 
Within the experimental and theoretical uncertainties this is fully consistent
with the present theory prediction, which is based on a perturbative 
NNLL QCD analysis \cite{Buras:2005gr}  and on a recent improvement 
of the long-distance contributions 
\cite{Isidori:2005xm}:
\begin{eqnarray} \nonumber
{\rm BR}(K^+\to\pi^+ \nu\bar{\nu})_{\rm SM} = \left( 0.80 \pm 0.11 \right) \times 10^{-10}~.
\end{eqnarray}
The error is dominated by parametrical errors on 
CKM matrix elements and on $m_c$ which can be significantly reduced in the future.

\begin{figure}[htbp]
\begin{center}
\vspace{-0.5cm}
\includegraphics[scale=0.42]{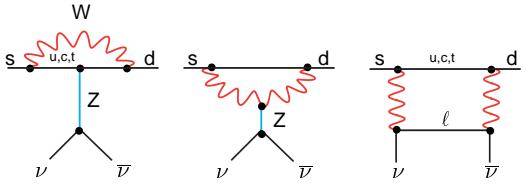}
 \vspace{-1cm}
\caption{ Graphs for $s\rightarrow d\nu\bar\nu$ in the SM
\label{SMkaon}}
\end{center}
\end{figure}

The rare  decays
$K^{+}\rightarrow\pi^{+}\nu\bar\nu$ and
$K_{L}\rightarrow\pi^{0}\nu\bar\nu$ are both exceptionally clean modes, 
  for two reasons essentially. First, the 
hard  (quadratic) GIM mechanism, 
is active; thus,
these decays are dominated by short-distance dynamics.
In fact, at the quark level the two processes arise from the 
$s\rightarrow d\nu\bar\nu$ process, which originates from 
a combination of the $Z$ penguin and a double $W$ exchange 
(see Fig.\ref{SMkaon}).

In these graphs the $u,c,t$ quarks appear as internal lines.
The hard GIM mechanism implies on the amplitude level:  
\begin{eqnarray} \nonumber 
A_q  \sim m_q^2/m_W^2  V^*_{qs} V_{qd},\,\, q=u,c,t\,. 
\end{eqnarray}
Thus, the top-quark contribution dominates, with a smaller
contribution, in the case of the
$K^{+}\rightarrow\pi^{+}\nu\bar\nu$ decay, from the charm contribution.
The up-quark contribution is in both cases negligible, so that
$s\rightarrow d\nu\bar\nu$ is essentially a short-distance
process. 

Moreover, the short-distance amplitude is governed 
by a single semileptonic operator whose
 hadronic  matrix element can be determined experimentally
by  the semileptonic kaon decay $K^+ \rightarrow \pi^0 e^+ \nu$ using isospin
symmetry; so the main hadronic uncertainties can be 
eliminated by experimental data.

\begin{figure}[htbp]
\begin{center}
\vspace{-0.5cm}
\includegraphics[scale=0.42]{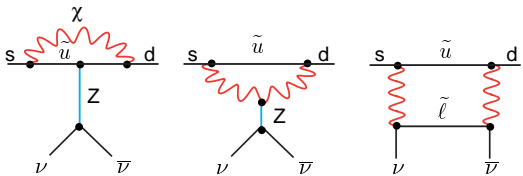}
 \vspace{-1cm}
\caption{ Graphs for $s\rightarrow d\nu\bar\nu$ in supersymmetry\label{MSSMkaon}}
\end{center}
\end{figure}

Besides their  rich CKM phenomenology,  the decays  
$K_L \rightarrow \pi^0\nu\bar{\nu}$ and 
$K^+ \rightarrow \pi^+ \nu \bar{\nu}$, as loop-induced processes, 
 are  very sensitive to new physics beyond the SM
and are crucial tools to discriminate between various new physics
scenarios in the future. 
Thanks to the cleanliness of the theoretical predictions, 
the measurement of these decays leads  to 
very accurate constraints on any new physics model and will
help us to discriminate between various new physics scenarios in the
future,  when new physics is discovered within the direct search.  
Moreover, there are also very interesting and theoretically 
clean correlations between 
$B$ and $K$ physics 
allowing for crucial precision tests of the SM and also of 
so-called minimal flavour violation scenarios  
in which the flavour structure  is essentially SM-like.
These correlations are generally violated in models with new sources 
of flavor violation.
There is also the possibility that these clean rare decay modes
themselves lead to direct evidence for new physics, if the measured 
decay rates are not compatible with the SM.
New effects in supersymmetric models,  for example, can be induced
through new box- and penguin-diagram contributions which involve
new particles such as charged Higgs or  charginos and stops 
(Fig.~\ref{MSSMkaon}), 
that replace the $W$ boson and the up-type quark  of the SM (Fig.~\ref{SMkaon}).Explicit analyses of possible post-SM scenarios,  
with direct new-physics contributions in the 
$s \rightarrow d \bar\nu\nu$  amplitude or in $B\bar B$ mixing, 
can be found   in~\cite{D'Ambrosio:2001zh} and \cite{Buras:2004uu}.

\begin{figure}[htbp]
\begin{center}
\includegraphics[scale=0.35]{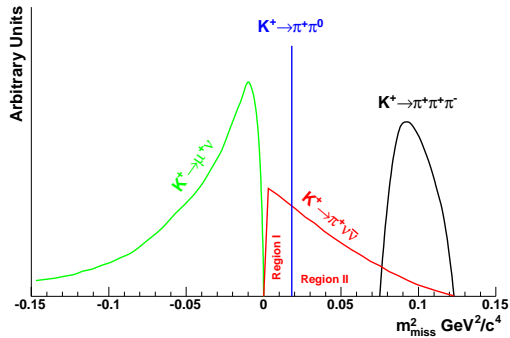}
 \vspace{-1cm}
\caption{ 
Distribution of the missing mass squared for the signal and the most
frequent kaon decays. The signal region is divided into region
I, which  lies between the two prominent two-body decays of the charged kaon,
while region II extends to the three-pion kinematical limit \cite{Augusto}.
\label{Exkaon}}
\end{center}
\end{figure}

Besides the recently finalized Brookhaven experiments on the charged mode
\cite{Adler:2004hp,Anisimovsky:2004hr}
and the running experiment on the neutral mode at KEK \cite{Inagaki},
there are two new proposals for the charged mode\cite{Komatsubara,Anelli},
 one for the neutral \cite{Hsiung}
currently  under discussion.

The novel feature of the proposed measurement of 
$K^+ \rightarrow \pi^+ \nu \bar \nu$ at CERN~\cite{Anelli} is 
that,   in contrast to previous experiments,  
one does not use stopped kaons but kaons in flight.  
One expects 80 $K^+ \rightarrow \pi^+ \nu \bar \nu$ events in about two years of data taking, starting in 2010, based on the SM predictions.

As can be seen in Fig.~\ref{Exkaon}, the main background is the mode
$K^+ \rightarrow \pi^+ \pi^0$. Thus, one of the 
main issues is the vetoing of the photons out of the $\pi^0$ decay, which is 
much easier than in previous experiments because of the high-energy kaon beam.

\subsection{Charmless rare $B$ decays: $B \rightarrow \pi\pi$}

Charmless rare $B$ decays  are an ideal testing ground for the QCD 
factorization approach. 
The QCD factorization theorems on non-leptonic decay modes, 
first proposed in \cite{Beneke:1999br}, 
identify short-distance effects ($T_{\rm I}$ and $T_{II}$) 
 that  can be systematically calculated in perturbation theory:
  $\langle PP| H_{\rm eff}|B\rangle
= F^{B \to P} \cdot T_{\rm I} \otimes \phi_P+T_{\rm II} \otimes \phi_B \otimes \phi_P \otimes \phi_P +  \mbox{terms suppressed by $1/m_b$}$,\,\,\,
where the symbol $\otimes$ represents convolution
with respect to the light-cone momentum fractions of
light quarks inside the mesons.
Non-perturbative effects are parametrized in terms of a few universal 
functions such as  form factors ($F^{B \to P}$) 
and light-cone distribution amplitudes ($\phi_B,\phi_P$), 
on which our information is rather restricted. 
Moreover, phenomenological applications are limited by the insufficient 
information on (power-suppressed) non-factorizable terms. 
Thus, the limiting  factor of the QCD factorization theorems is twofold, 
namely the insufficient information on their  non-perturbative input and 
on the power-suppressed (non-factorizable) terms. 
This problem should be 
tackled by the large data sets made available by current  and future 
$B$ physics experiments and by the development of
 suitable new non-perturbative methods on the theory side.

\begin{figure}[htbp]
\begin{center}
\vspace{-0.5cm}
\includegraphics[scale=0.43]{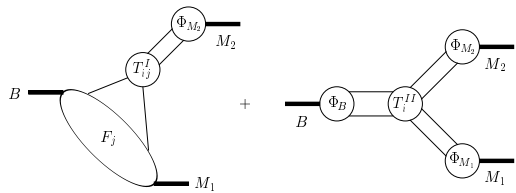}
\vspace{-0.7cm}
\caption{Illustration of QCD factorization theorems for non-leptonic 
 $B$ decays}
\end{center}
\label{factorization}
\end{figure}

If one does not want to rely on particular assumptions about the
hadronic dynamics, the traditional model-independent approach to
charmless non-leptonic $B$\/-decays is to decompose the various
decay amplitudes according to isospin  \cite{Gronau:1990ka,Nir:1991cu}.
The present experimental data on $B \to \pi\pi$ decays is consistent
with this isospin decomposition. 
In \cite{Feldmann:2004mg} it was proposed to use the
model-independent predictions from QCD factorization for the 
factorizable part of the decay amplitudes. The standard isospin analysis
is then applied to the non-factorizable contributions only, which thus
covers the model-dependent estimate of chirally enhanced power corrections,
large deviations from the standard hadronic input parameters, etc. 
Comparing with the experimental data on branching ratios and 
CP~asymmetries, this method allows to quantify
the amount of non-factorizable effects in particular isospin
amplitudes without additional theoretical bias. 

From the present data on $B \to \pi\pi$ decays, 
one finds that small non-factorizable contributions are disfavoured.  
Clearly, the dynamical origin of these non-factorizable corrections 
remains a theoretical challenge. At present, 
different phenomenological assumptions can accommodate the data.  
For instance, the latest update \cite{Beneke:2003zv} of the BBNS approach
\cite{Beneke:2001ev} advocates scenarios where certain
hadronic input parameters are put to the edges of the
allowed regions and the estimate about the size of a particular class of 
$1/m_b$ corrections is included in the theoretical uncertainties.
Other authors assume the dynamical enhancement of certain
flavour topologies to explain the experimental data, 
for instance the charming penguin approach of  
\cite{Ciuchini:2001gv}. Also a new analysis \mbox{using} SCET methods was
recently presented \cite{Bauer:2005kd}, where SCET relations in 
leading order are 
combined with $SU(3)$ flavour relations. This leads to a further reduction 
of hadronic parameters due to the vanishing of some  strong phases 
in the $m_b \rightarrow \infty$ limit.

\begin{figure}[htbp]
\begin{center}
\vspace{-0.5cm}
\includegraphics[scale=0.6]{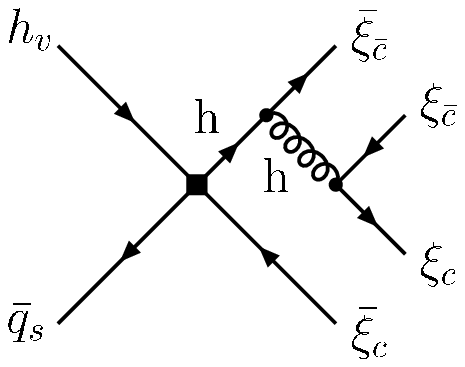}\,\,\,\, \includegraphics[scale=0.6]{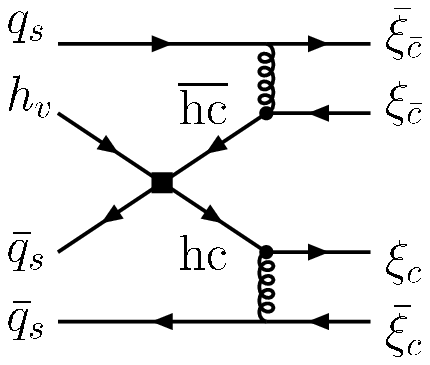}
\vspace{-0.7cm}
\caption{Annihilation contributions to $B \to PP$. 
  (a) Example for one-gluon exchange
   included in the
   BBNS analysis. (b) Example for higher Fock-state-contribution
 \label{annih}}
 \end{center}
 \end{figure}

Often only dominant $1/m_b$ corrections are identified corresponding 
to simple $1/m_b$ SCET operators. However,  
the $1/m_b$ corrections also include more power-suppressed decay 
current operators and 
more insertions of subleading Lagrangian terms which lead to 
terms sensitive to higher Fock-state-contributions (see Fig.\ref{annih})
which are not necessarily suppressed. 

Finally, it should be stressed that any dynamical
assumption which further constrains the isospin analysis 
may lead to a strong bias when used in CKM fits. Therefore, one
should rather use the experimental data itself to distinguish
between different alternatives, and to measure the power-suppressed
non-factorizable matrix elements in QCD factorization/SCET.

\subsection{Is there a $B \rightarrow K \pi$ puzzle?}

The $B \rightarrow K \pi$ modes are well-known for  being  sensitive  
to new electroweak penguins beyond the SM 
\cite{Fleischer:1997ng,Grossman:1999av}.
The present data on CP averaged $K\pi$ branching ratios can be expressed in 
terms of three ratios:
\begin{eqnarray}
{R}=\frac{\tau_{B^+}}{\tau_{B^0}}
    \frac{{\rm BR}[B_d^0 \to \pi^-K^+]+{\rm BR}[\bar B_d^0 \to \pi^+K^-]}
        {{\rm BR}[B_d^+ \to \pi^+K^0]+{\rm BR}[B_d^- \to \pi^- \bar K^0]}\nonumber\\
 {R_n}=\frac12 \frac{{\rm BR}[B_d^0 \to \pi^-K^+]+{\rm BR}[\bar B_d^0 \to \pi^+K^-]}
        {{\rm BR}[B_d^0 \to \pi^0K^0]+{\rm BR}[\bar B_d^0 \to \pi^0 \bar K^0]}\nonumber\\
 {R_c}=2\frac{{\rm BR}[B_d^+ \to \pi^0 K^+]+{\rm BR}[B_d^- \to \pi^0 K^-]}
        {{\rm BR}[B_d^+ \to \pi^+ K^0]+{\rm BR}[B_d^- \to \pi^- \bar K^0]}\nonumber
\end{eqnarray}
The present data \cite{HFAG},  
\begin{eqnarray}\nonumber
R=0.84^{+0.06}_{-0.05},R_n=0.82^{+0.08}_{-0.07},R_c=1.00^{+0.09}_{-0.08},
\end{eqnarray}
appears somewhat anomalous, 
when compared, for example, with the \mbox{approximate} sum rule 
proposed in \cite{Lipkin:1998ie,Gronau:1998ep,Matias:2001ch}            
which leads to the prediction  $R_c = R_n$.
The corresponding BBNS predictions \cite{Beneke:2003zv,BBNSprivate} are
\begin{eqnarray} \nonumber
R=0.91^{+0.13}_{-0.11}, R_n=1.16^{+0.22}_{-0.19}, R_c=1.15^{+0.19}_{-0.17}.
\end{eqnarray}
One can also use approximate flavour
symmetries (isospin or $SU(3)_F$) to relate different
decay amplitudes and reduce the number of unknown hadronic
parameters \cite{Gronau:1990ka,Nir:1991cu}.
This procedure is often 
combined with additional dynamical assumptions about
the importance of certain flavour topologies that can
be identified in the factorization approximation only. 
In particular, in 
a recent study \cite{Buras:2004ub} along these lines, 
the $B \rightarrow \pi\pi$ data was used to make theoretical
predictions on the $B \rightarrow K \pi$ modes. This specific 
approach   leads to 
\begin{eqnarray} \nonumber
R=0.94^{+0.03}_{-0.03}, R_n=1.14^{+0.08}_{-0.07}, R_c=1.11^{+0.06}_{-0.07}.
\end{eqnarray} 
At present, it is difficult to estimate the theoretical uncertainties due 
to the assumptions made in this
approach, but in the future those assumptions including the $SU(3)_F$ 
symmetry can be tested experimentally.

Does this slightly anomalous behaviour guide us to new physics in the $K \pi$ 
modes? 
It is important to note that 
the deviation of $R_n$ and $R_c$ from unity is solely due to isospin-breaking
effects \cite{Matias:2001ch} and 
the amount of short-distance isospin breaking in the
SM is too small to explain the experimental number.
Whereas the authors of  \cite{Buras:2004ub} 
argue that this may point
to an interesting avenue towards new physics in electroweak
penguin operators, the collaboration in \cite{Charles:2004jd}
considers this deviation as a statistical fluctuation, which
is consistent with the SM -- 
even when the $SU(3)_F$ constraints are enforced.  
Not surprisingly, the $B \to K \pi$ data has 
triggered several new-physics analyses.

Due to the large non-factorizable contributions identified in
the $B \to \pi\pi$ channel, it is difficult to decide 
whether the data on $B \to \pi K$ decays point to new physics 
(for example to isospin breaking via new 
degrees of freedom as discussed in \cite{Fleischer:1997ng,Grossman:1999av}), 
or whether they can be explained by non-factorizable 
$SU(3)_F$- or isospin-violating QCD and QED effects within the SM.
For the latter possibility, it was pointed out in
\cite{Feldmann:2004mg} that the actual expansion parameter for
non-perturbative isospin violating effects in $B\to \pi K$
can be as large as $5-10\%$, for instance via non-factorizable 
corrections from CKM\/-enhanced long-distance QED penguins.
This seems to be in the right ball park 
to at least partly explain the deviation of $R_n$ from unity.

Apart from the theoretical questions about the interpretation of the
data, there are also some experimental issues:

\begin{itemize}
\item
Radiative corrections to decays with charged particles in the final
states had not been taken into account properly
in previous analyses, an effect which was expected to lead 
to an increased branching ratio of these modes (see \cite{Charles:2004jd})
and which could bring $R_n$ closer to unity. 
In fact, quite recently
such effects were calculated \cite{Baracchini:2005wp} and found to be 
not negligible. On the experimental side, BABAR has already taken into 
account these results on the final state interactions in order to 
estimate the efficiency by comparing with the used  Monte Carlo program 
PHOTOS  \cite{Barberio:1993qi}. 
 The new experimental branching ratios quoted by BABAR correspond now to a
 well-defined photon energy-cut. These corrections on the experimental side already led to a $+6\%$ correction  in the $K^+\pi^-$ mode and $+8\%$ in the
$\pi^+\pi^-$ mode \cite{Aubert:2005ni}. 
However, the theoretical prediction still has to be corrected accordingly 
in order to be compared with the experimental data using a photon energy-cut. 
There is an additional effect of around $3-5\%$ expected 
\cite{Baracchini:2005wp}, but the precise value depends 
on the matching with the short-distance calculation. 
\item   It has also been argued in \cite{Gronau:2003kj}
that the present pattern 
could perhaps result from underestimating the $\pi^0$ detection efficiency 
which implies an overestimate for any branching ratio involving a $\pi^0$.
The authors of \cite{Gronau:2003kj}
propose therefore to consider the 
ratio $(R_n R_c)^{1/2}$ in which the $\pi^0$ detection efficiency
cancels out. 
However, this  idea has not found any confirmation on the experimental side. 
\end{itemize}

One  may also shed some light on the
dynamical origin of non-factorizable effects, which
may stimulate further studies with non-perturbative methods.
In particular, a systematic classification of power-suppressed
matrix elements from non-factorizable SCET operators should
give an alternative scheme compared to the traditional
classification in terms of flavour topologies.

\section{Search  in CP-violating observables}
The CKM prescription of CP violation with a single phase is very
predictive. It was proposed already in 1973~\cite{CKM72}, before the
experimental confirmation of the existence of the second family.
Before the start of the $B$ factories, the neutral kaon system was the
only environment where CP violation had been observed.  It has been
difficult to decide if the CKM description of the SM
really accounted quantitatively for the CP violation observed in the
kaon system, because of the large theoretical uncertainties due to
long-range strong interactions. The rich data sets from the
$B$ factories now allow for an independent and really quantitative
test of the CKM-induced CP-violating effects in several independent
channels. Within the golden $B$ mode, $B_d \rightarrow J/\psi K_S$, the
CKM prescription of CP violation has already passed its first
precision test; in fact, the measured CP violation is well in
agreement with the CKM prediction~\cite{Belle,Babar}.
\begin{figure}[htbp]
\begin{center}
\vspace{-0.5cm}
\includegraphics[scale=0.4]{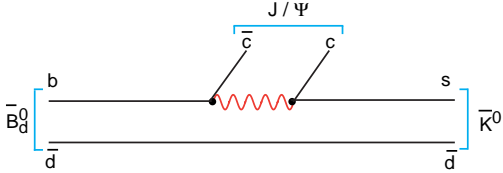}
 \vspace{-1cm}
\caption{The tree-level process  $B_d \rightarrow J/\psi K_S$}
\end{center}
\label{CPtree}
\end{figure}
Nevertheless, there is still room for non-standard CP phases. An
additional experimental test of the CKM mechanism is provided by the
mode $B_d \rightarrow \Phi K_S$. This mode is induced at the loop
level only and, therefore, it is much more sensitive to possible
additional sources of CP violation than the tree-level-induced decay
$B_d \rightarrow J/\psi K_S$. However, the poor statistics does not
allow to draw final conclusions yet~\cite{Belle,Babar}. 
We also note that the commonly used average over various 
$b \rightarrow s\bar s s$ modes are not a reasonable procedure because of
 the  different hadronic uncertainties  of those modes. 
\begin{figure}[htbp]
\begin{center}
\vspace{-0.5cm}
\includegraphics[scale=0.4]{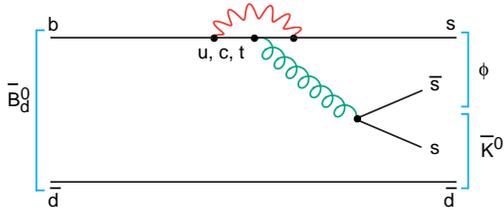}
 \vspace{-1cm}
\caption{The loop-induced process  $B_d \rightarrow \Phi K_S$}
\end{center}
\label{CPloop}
\end{figure}
There are 
also direct CP asymmetries in loop-induced $\Delta F = 1$ modes,  
which allow for
additional precision tests of the mechanism of CP violation 
\cite{Hurth:2001yb,Hurth:2001ja,Hurth:2003dk}.
Currently, these decays are less probed than $\Delta F =2$
transitions. However, very precise measurements of direct CP
asymmetries in inclusive rare $B$ decays, such as $b \rightarrow s$ or
$b \rightarrow d$ transitions, will be possible in the near future.
A first measurement of the so-called untagged direct CP violation
in the summed inclusive $b \rightarrow s+d \gamma$ was already
presented by the BABAR collaboration \cite {Aubert:2005cb}.

\section{Interplay of flavour and collider physics}
Within the next decade an important interplay 
of flavour and collider physics 
will take place. As from the day the existence of new degrees 
of freedom  is established  by  the LHC, the study of  anomalous  phenomena  
in the flavour sector will  become an important tool for studying their 
phenomenology.  

For example, within supersymmetric extensions of the SM, the 
measurement of the flavour structure is directly linked to the crucial 
question of the supersymmetry-breaking mechanism. Thus, the 
flavour sector is  important in  distinguishing  between models
of supersymmetry. 
This example nicely shows the obvious complementary nature of flavour physics 
and  collider  physics.  At the LHC    
direct searches for supersymmetric particles are  essential in establishing  
the existence of new physics. 
On the other hand, there are a variety of possibilities for the origin of 
flavour structures within  supersymmetry. Flavour physics 
provides an important 
tool  with which fundamental questions, such as how supersymmetry is broken, 
can be addressed.

Until now the focus within the direct search for supersymmetry has been 
mainly on flavour diagonal observables. But there are 
many obvious  relations  
between non-diagonal collider and 
low-energy flavour observables. In \cite{Hurth:2003th}, 
flavour violation in the 
squark decays of the second and third generations, were analysed, taking  
into account results from $B$ physics, in particular from the rare decay 
$b \to s \gamma$.
It was  shown that correlations between various squark decay modes can be 
used  to obtain more precise information on various flavour-violating 
parameters.

Within the minimal supersymmetric standard model (MSSM),  
 there are two new sources of FCNCs, namely new contributions, which are 
induced through the quark mixing as in the SM and generic supersymmetric 
contributions through  the squark  mixing. The 
 structure of the MSSM 
does not explain the suppression of FCNC processes which is observed 
in experiments (supersymmetric flavour problem).
The generic new  sources of flavour violation that may be present in
supersymmetric models, in addition to those enclosed in the CKM matrix,
are a direct consequence of a possible disalignment of quarks and squarks.
One has to consider the various  contributions to the squark mass matrices
\begin{eqnarray}
\left(\begin{array}{cc}
  M^2_{f,LL}+F/D_{f,LL} & M_{f,LR}^2 + F_{fLR}\\ 
\left(M_{f,LR}^{2}\right)^{\dagger}+F_{f,RL} & M^2_{f,RR}+F/D_{f,RR}     
\end{array} \right) \nonumber 
\label{massmatrixd}
\end{eqnarray}
where $f$ stands for up- or down-type squarks.
The matrices $M_{u,LL}$ and $M_{d,LL}$  are related 
by $SU(2)_L$ gauge invariance. In the super-CKM basis, 
where the quark mass matrices are diagonal 
and the squarks are rotated in parallel to their superpartners, the relation
reads as 
$K^\dagger M^2_{u,LL} K =  M^2_{d,LL} = M^2_Q$. In this basis the $F$-terms
$F_{f\,LL}$, $F_{f\,RL}$, $F_{f\,RR}$, as well as the $D$-terms $D_{f\,LL}$
and  $D_{f\,RR}$, are diagonal. All the additional flavour structure of the
squark sector is encoded in the soft SUSY-breaking terms $M^2_Q$,
$M^2_{\,f,\,RR}$                                             and 
$M_{\,f,\,LR}^2$.
 These additional
flavour structures induce flavour-violating couplings to the neutral
gauginos and higgsinos in the mass eigenbasis, which give rise to
additional contributions to observables in the $K$- and $B$-meson sector.

At present, new physics contributions to $s \to d$ and $b\to d$ 
transitions are strongly constrained.
In particular,  the transitions between first- and second-generation 
quarks, namely FCNC processes in  the $K$ system, are 
the most formidable tools to shape viable supersymmetric 
flavour models. Nevertheless, most of the phenomena involving  
$b \rightarrow s$ transitions  are still largely unexplored and 
leave open the possibility of large 
new physics effects, in spite of the strong bound of the
famous $\bar B \rightarrow X_s \gamma$ decay,  which 
still gives the most stringent bounds in this sector. 
 Nevertheless, additional experimental 
information from the $\bar B \rightarrow X_s \ell^+ \ell^-$ decay at the
$B$ factories and new results on the $B_s$--$\bar B_s$ 
mixing  at the Tevatron might change this situation in the near future.

Squarks can decay into quarks of all generations if the
most general form of he squark mass matrix is considered. The 
most important decay modes for the example under study are:
\begin{eqnarray}
\tilde u_i &\to& u_j \tilde \chi^0_k \, , \,  d_j \tilde \chi^+_l \nonumber\\
\tilde d_i &\to& d_j \tilde \chi^0_k \, , \,  u_j \tilde \chi^-_l \nonumber
\end{eqnarray}
with $i=1,..,6$, $j=1,2,3$, $k=1,..,4$ and $l=1,2$.

These decays are controlled by the same mixing matrices as the
contributions to $b \to s \gamma$. As this decay mode restricts
the size of some of the elements, the question arises regarding  
to which extent
flavour-violating squark decays are also restricted. The following tables 
show    that flavour-violating decay modes are hardly constrained 
by current data. 
We used the so-called Snowmass point SPS\#1a \cite{Allanach:2002nj}
as a specific example, 
which is characterized by $m_0=100$~GeV, $m_{1/2}=250$~GeV, $A_0=-100$~GeV,
$\tan\beta=10$ and $\rm{sign}(\mu)=1$. At the electroweak scale
one gets the following data: $M_2=192$~GeV, $\mu=351$~GeV, $m_{H^+}=396$~GeV,
$m_{\tilde g}=594$~GeV, $m_{\tilde t_1}=400$~GeV, $m_{\tilde t_2}=590$~GeV,
$m_{\tilde q_R} \simeq 550$~GeV, and $m_{\tilde q_L} \simeq 570$~GeV.
We concentrated on the mixing between the second
and third generations and we used  
one point in the parameter space representing 
typical results  fulfilling the $b\to s \gamma$ constraint based on a LL
QCD analysis as given in \cite{Borzumati:1999qt,Besmer:2001cj}.
In  Tables~\ref{tab:br} and  \ref{tab:brsd}  we have collected
the branching ratios of squarks that are larger than 1\%.
Clearly, all considered particles have large
flavour-changing decay modes.

\begin{table*}
\caption{Typical branching ratios (in \%) of $u$-type squarks}
\label{tab:br}
\begin{center}
\begin{tabular}{|c|cc|cc|cc|cc|cc|cc|}
\hline
          & $\tilde \chi^0_1 c$ &  $\tilde \chi^0_1 t$ & 
              $\tilde \chi^0_2 c$ &  $\tilde \chi^0_2 t$ & 
              $\tilde \chi^0_3 c$ &  $\tilde \chi^0_3 t$ &
              $\tilde \chi^0_4 c$ &  $\tilde \chi^0_4 t$ &
              $\tilde \chi^+_1 s$ &  $\tilde \chi^+_1 b$ & 
              $\tilde \chi^+_2 s$ &  $\tilde \chi^+_2 b$ \\ \hline 
$\tilde u_1$ & 4.7                & 18 &
               5.2                &  9.6 & 
               $6 \times  10^{-3}$  & 0    &
               0.02               & 0    &
              11.3                & 46.4   &
               $2 \times  10^{-3}$  & 4.7 \\  
$\tilde u_2$ & 19.6               & 1.1  &
               0.4                & 17.5 &
               $2 \times 10^{-2} $ & 0  &
               $6 \times 10^{-2} $ & 0 &
               0.5                & 57.5 &
               $3 \times  10^{-3} $ &  2.9 \\
$\tilde u_3$ & 7.3                & 3.7  &
               20                 & 1.4  &
              $6 \times 10^{-2} $ & 0    &
               0.6                & 0    &
               40.3               & 3.1  &
               1                  & 18.5 \\
$\tilde u_6$ & 5.7                &  0.4  &
              11.1                &  5.3  &
               $4 \times 10^{-2} $ &  5.7  &
              0.6                 & 13.2  &
             22.9                 & 13.1  &
             0.6                  &  8.0 \\ 
\hline
\end{tabular}
\end{center}
\end{table*}

\begin{table*}
\caption{Typical branching ratios (in \%) of $d$-type squarks} 
\label{tab:brsd}
\begin{center}
\begin{tabular}{|c|cc|cc|cc|cc|cc|cc|c|}
\hline
          & $\tilde \chi^0_1 s$ &  $\tilde \chi^0_1 b$ & 
              $\tilde \chi^0_2 s$ &  $\tilde \chi^0_2 b$ & 
              $\tilde \chi^0_3 s$ &  $\tilde \chi^0_3 b$ &
              $\tilde \chi^0_4 s$ &  $\tilde \chi^0_4 b$ &
              $\tilde \chi^-_1 c$ &  $\tilde \chi^-_1 t$ & 
              $\tilde \chi^-_2 c$ &  $\tilde \chi^-_2 t$ &
              $\tilde u_1 W^-$ \\ \hline 
$\tilde d_1$ & 1.2                & 5.7 &
               8.4                &  30.6 & 
               $2 \times 10^{-2}$  & 1.5    &
               0.2               & 0.9    &
              16.6                & 34.1   &
               0.6  & 0 & 0 \\  
$\tilde d_2$ & 17.4               & 5.8  &
               5.1                & 15.7 &
               $7 \times 10^{-2} $ & 7.4  &
               0.3 & 09.2 &
               9.7                & 19.7 &
               0.7 &  0 & 8.8 \\
$\tilde d_4$ & 14.7                & 21.7  &
               11.3                & 2.2  &
              $5 \times 10^{-2} $ & 10.6    &
               0.5                & 8.4    &
               22.1               & 3.6  &
               1.2                  & 0 & 3.4\\
$\tilde d_6$ & 1.7                &  0.5  &
              20.5                &  6.9  &
               0.1 &  0.9  &
              1.2                 & 1.3  &
             40.3                 & 10.2  &
             3.4                  & 11.1 & 1.8 \\ 
\hline
\end{tabular}
\end{center}
\end{table*}
The results demonstrate  that large flavour-changing decays of squarks
and gluinos  are consistent with current data from the Tevatron and the $B$ 
factories. Such a situation makes life at the 
LHC potentially more interesting, but
more difficult. It is clear that it would  be complicated  or even impossible
to disentangle the various decay modes at the LHC;   
 flavour tagging is not sufficient. 
In these cases, additional information from 
future flavour experiments will be necessary  to interpret
that  LHC data properly.

\section{Outlook}
It is expected that the experiments at the LHC will 
lead to discoveries of new  degrees  of freedom at the TeV energy scale. 
The precise nature of this new physics 
is unknown, but it is strongly expected  that it will  answer 
some of the fundamental questions related to the origin 
of electroweak symmetry breaking.
Independently of the nature of the new  physics, a flavour physics
program   
parallel to the LHC's will be crucial to disentangle 
the precise  features of the newly uncovered phenomena and 
to discriminate between 
different new physics scenarios. 
In particular, the measurement 
of theoretically clean loop-induced rare $B$- and $K$-meson 
decays  as   highly sensitive probes for new degrees of freedom 
beyond the SM will lead to important information complementary to collider 
data;
there are important fundamental questions that will be 
addressed exclusively  by future flavour experiments, for \mbox{example}
how FCNCs  are suppressed beyond the SM 
(flavour problem), if  there exist 
new sources of flavour and CP violation beyond
those in the SM, if there is  CP violation in the QCD gauge sector, 
how neutrino masses are generated, and what the
relation between the flavour structure in the lepton and quark
 sectors is. 
All these questions include exciting options to learn something about 
physics at a scale  much higher than our current  experiments. 
Thus, a diversified and thorough experimental programme  in flavour physics 
will continue to be an essential element for the understanding of nature.\\

{\bf Acknowledgement}\\  
I would like to thank Augusto Ceccucci, Markus Cristinziani, 
Carlo Dallapiccola, Jerome Charles, and Gino Isidori for useful discussions.  
I am also very grateful to Thorsten Feldmann 
for a careful and critical reading of the manuscript.

\end{document}